\documentclass{article}[10pt]

\usepackage{epsfig}

\def\encadre#1#2{%
\setbox100=\hbox{\kern#1{#2}\kern#1}
\dimen100=\ht100 \advance \dimen100 by #1
\dimen101=\dp100 \advance \dimen101 by #1
\setbox100=\hbox{\vrule height \dimen100 depth \dimen101\box100\vrule}
\setbox100=\vbox{\hrule\box100\hrule}
\advance \dimen100 by .4pt \ht100=\dimen100
\advance \dimen101 by .4pt \dp100=\dimen101
\box100
\relax
}

\def\leurre{\noindent\leftskip0pt\small\baselineskip 10pt}

\newtheorem{thm}{Theorem}

\newtheorem{fig}{Figure}

\newcount\figno\figno=1
\newcount\tabno\tabno=1
\def\legende taille #1 Leg #2 {%
               \setbox148=\hbox{\rmix Figure\ \the\figno.}
               \setbox149=\vtop{\parindent0pt\baselineskip 10pt
                                \hsize=#1
                                \rmix\mathix
                                 #2}
                    \ligne{\hfill \box148\hskip 7pt\box149\hfill}
               \global\advance\figno by 1}
\def\legendetab taille #1 Leg #2 {%
               \setbox148=\hbox{\rmix Table\ \the\tabno.}
               \setbox149=\vtop{\parindent0pt\baselineskip 10pt
                                \hsize=#1
                                \rmix\mathix
                                 #2}
                    \ligne{\hfill \box148\hskip 7pt\box149\hfill}
               \global\advance\tabno by 1}
\newcount\nbtheo\nbtheo=1
\newcount\nbdefi\nbdefi=1
\newcount\nbcorol\nbcorol=1
\newcount\nblemma\nblemma=1
\newcount\nbpropos\nbpropos=1
\def\novtheo{\vskip 0pt\noindent
             {\bf Theorem~\the\nbtheo}{}\global\advance\nbtheo by 1}
\def\novdefi{\vskip 0pt\noindent
             {\bf Definition~\the\nbdefi}{}\global\advance\nbdefi by 1}
\def\novcorol{\vskip 0pt\noindent
             {\bf Corollary~\the\nbcorol}{}\global\advance\nbcorol by 1}
\def\novlemma{\vskip 0pt\noindent
              {\bf Lemma~\the\nblemma}{}\global\advance\nblemma by 1}
\def\novpropos{\vskip 0pt\noindent
              {\bf Proposition~\the\nbpropos}{}\global\advance\nbpropos by 1}

\def\cqfd{\hbox{\kern 2pt\vrule height 6pt depth 2pt width 8pt\kern 1pt}}
\begin{document}
\def\ligne#1{\hbox to \hsize{#1}}
\def\PlacerEn#1 #2 #3 {\rlap{\kern#1\raise#2\hbox{#3}}}
\font\bfxiv=cmbx12 at 12pt
\font\bfxii=cmbx12
\font\bfix=cmbx9
\font\bfvii=cmbx7
\font\bfvi=cmbx6
\font\bfviii=cmbx8
\font\pc=cmcsc10
\font\itix=cmti9
\font\itviii=cmti8
\font\rmix=cmr9 \font\mmix=cmmi9 \font\symix=cmsy9
\def\mathix{\textfont0=\rmix \textfont1=\mmix \textfont2=\symix}
\font\rmviii=cmr8 \font\mmviii=cmmi8 \font\symviii=cmsy8
\def\mathviii{\textfont0=\rmviii \textfont1=\mmviii \textfont2=\symviii}
\font\rmvii=cmr7 \font\mmvii=cmmi7 \font\symvii=cmsy7
\def\mathvii{\textfont0=\rmvii \textfont1=\mmvii \textfont2=\symvii}
\font\rmvi=cmr6 \font\mmvi=cmmi6 \font\symvi=cmsy6
\def\mathvi{\textfont0=\rmvi \textfont1=\mmvi \textfont2=\symvi}
\font\rmv=cmr5 \font\mmv=cmmi5 \font\symv=cmsy5
\def\mathv{\textfont0=\rmv \textfont1=\mmv \textfont2=\symv}
\font\rmxii=cmr12 \font\mmxii=cmmi12 \font\symxii=cmsy12
\def\mathxii{\textfont0=\rmxii \textfont1=\mmxii \textfont2=\symxii}
\font\rmxiv=cmr14 \font\mmxiv=cmmi14 \font\symxiv=cmsy14
\def\mathxiv{\textfont0=\rmxiv \textfont1=\mmxiv \textfont2=\symxiv}
\font\rmvii=cmr7
\font\rmv=cmr5

\newcount\figdecompte
\global\figdecompte=1
\def\novfignum #1 {\newcount#1
                    \global#1=\figdecompte
                    \global\advance\figdecompte by 1}
\novfignum{\figun}
\novfignum{\tilvqiv}
\novfignum{\splitvqiv}
\novfignum{\splitspanvqiv}
\novfignum{\numspanvqiv}
\novfignum{\splitpqevz}
\novfignum{\splitpqevo}
\novfignum{\splitpqodz}
\novfignum{\splitpqodo}
\novfignum{\rulehiii}
\novfignum{\splithiii}
\novfignum{\rulehii}
\novfignum{\splithii}
\novfignum{\algopq}
\novfignum{\elemcycpqa}
\novfignum{\elemcycpqb}

\newcount\theodecompte
\global\theodecompte=1
\def\novtheonum #1 {\newcount#1
                    \global#1=\theodecompte
                    \global\advance\theodecompte by 1}
%\novtheonum {\combpqiv}
%\novtheonum {\regpqiv}
\novtheonum {\preferred}
%\novtheonum {\status}
%\novtheonum {\father}
%\novtheonum {\linalgo}
%\novtheonum {\linpath}
%\novtheonum {\combviiqiii}
%\novtheonum {\spanpqivqpiiqiii}
\novtheonum {\combpqq}
\novtheonum {\hollander}
\novtheonum {\spanpqq}
\novtheonum {\linpqq}

\newcount\lemmadecompte
\global\lemmadecompte=51
\def\novlemmanum #1 {\countdef#1=\lemmadecompte
                    \advance\lemmadecompte by -50
                    \global#1=\lemmadecompte
                    \advance\lemmadecompte by 50
                    \global\advance\lemmadecompte by 1}
\novlemmanum {\forbidfibo}
\novlemmanum {\auxilforbid}
\novlemmanum {\ruleprefqv}
\novlemmanum {\ukmini}
\novlemmanum {\localemm}
\novlemmanum {\locabeta}
\novlemmanum {\perphalfplane}
\novlemmanum {\separvqiv}
\novlemmanum {\separviiqiii}
\novlemmanum {\convballs}
\novlemmanum {\borderball}

\newcount\corodecompte
\global\corodecompte=1
\def\novcoronum #1 {\newcount#1
                    \global#1=\corodecompte
                    \global\advance\corodecompte by 1}
\novcoronum {\cregpqiv}
\novcoronum {\ccombregpiiqiii}
\novcoronum {\cbigUn}
\novcoronum {\cposifather}
\novcoronum {\czzassignpref}
\novcoronum {\caleaballs}

\newcount\tabledecompte
\global\tabledecompte=1
\def\novtablenum #1 {\newcount#1
                    \global#1=\tabledecompte
                    \global\advance\tabledecompte by 1}

\novtablenum {\neighpqiv}
\novtablenum {\computpath}
\novtablenum {\rulehiv}
\novtablenum {\rulegene}
\novtablenum {\characonfig}
\novtablenum {\neighsons}
\novtablenum {\neighnotsons}

\def\ttV{\vrule depth 6pt height 12pt width 0.6pt}
\def\ttH{\hrule depth 0.3pt height 0.3pt width \hsize}

\title{About a new splittings for the algorithmic study of\\ the tilings $\{p,q\}$ 
of the hyperbolic plane\\ when $q$ is odd}

\author{{\bfseries Maurice Margenstern}\\
(Laboratoire d'Informatique Th\'eorique et Appliqu\'ee, EA 3097,\\
Universit\'e de Metz, \\
UFR-MIM, D\'epartement d'Informatique,\\
\^Ile du Saulcy,\\
57045 Metz Cedex, France,\\
margens@univ-metz.fr)\\
}
\maketitle

\pagenumbering{arabic}

\begin{abstract}
   In this paper, we remind previous results about the tilings $\{p,q\}$ of
the hyperbolic plane. We introduce two new ways to split the hyperbolic plane in order
to algorithmically construct the tilings $\{p,q\}$ when $q$ is odd.
\end{abstract}

%\begin{keywords}

%\begin{category}
%C.4.4.
%\end{category}
\section{Introduction}\label{intro}

   As mentioned in the abstract, the goal of this paper is to introduce two
ways within the algorithmic approach to the study of the tilings $\{p,q\}$ when $q$~is odd.
This approach is in the spirit of what has be done by the author starting from
the basic paper~\cite{mmJUCSii}.

   In Section~\ref{hypgeom}, we remind the reader the basic features of hyperbolic
geometry, in particular what is needed to define and to study the tilings $\{p,q\}$.
In Section~\ref{whenevenq}, we remind the splitting of the hyperbolic plane
leading to a spanning tree of the tiling $\{p,q\}$ when $q$~is even.
In Section~\ref{whenoddq}, we consider the tiling $\{p,q\}$ when $q$~is odd.
In Subsection~\ref{basic}, we remind the construction performed in~\cite{mmbook1}.
In Subsections~\ref{split1} and~\ref{split2}, we present the two ways of splitting
mentioned in the abstract. Both ways have nice properties except in one case,
when $p=4$ and $q=5$. Subsection~\ref{til45} is devoted to the study of this case.

\section{Tessellations in the hyperbolic plane}\label{hypgeom}

   In this section, we first remind Poincar\'e's disc model and a few features
which will allow us to define tessellations in the hyperbolic plane which we shall
study in Subsection~\ref{tessellations}
   
\subsection{Hyperbolic geometry}

   Hyperbolic geometry appeared in the first half of the 
19$^{\rm th}$ century, proving the independence of the parallel
axiom of Euclidean geometry. Models were devised in the second
half of the 19$^{\rm th}$ century and we shall use here one of the
most popular ones, Poincar\'e's disc. Figure~\ref{poincare_disc}
illustrates the parallel axiom of hyperbolic geometry in this model.
\vskip 7pt
\setbox110=\hbox{\epsfig{file=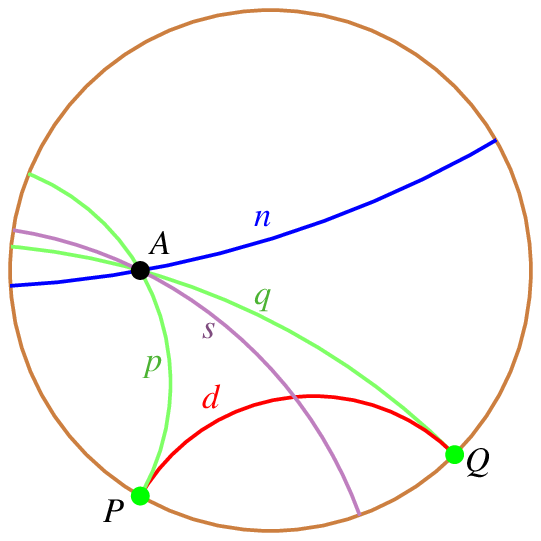,width=200pt}}
\vtop{
\ligne{\hfill
\PlacerEn {-100pt} {0pt} \box110
\hfill}
\vspace{-20pt}
\begin{fig}
\label{poincare_disc}\small
Illustration of the parallel axiom of hyperbolic geometry. The lines $p$ and~$q$
are the parallels to~$m$ which pass through~$A$.
\end{fig}
}

In Poincar\'e's disc model, the points of the hyperbolic plane are exactly those which
are inside a once for all fixed open disc of the Euclidean plane. The border of 
the disc is called the set of {\bf points at infinity}. Note that the points at
infinity do not belong to the hyperbolic plane. Lines are trace of diameters or circles 
orthogonal to the border of the disc, see the line~$m$ in Figure~\ref{poincare_disc}. 
In this figure, we have a point~$A$ not on~$m$ and we can see a line~$s$ which cuts~$m$. 
Two lines passing through~$A$ play a particular role: the lines $p$ and~$q$, which 
touch~$m$ in the model at~$P$ and~$Q$ which are points at infinity. These
lines are called {\bf parallel} to~$m$. More generally, in the model, two lines are
parallel if and only if they have a common point at infinity.

   Now, Figure~\ref{poincare_disc} shows that there is another important case
which has no counterpart in the Euclidean plane: the line~$n$ also passes 
through~$A$ without cutting~$m$, neither inside the disc nor outside it. Such 
a line is called {\bf non-secant} with~$m$. Non-secant lines are characterized by
the fact that they have a unique common perpendicular. This is also a specific property
of the hyperbolic plane where there cannot be rectangles. Another important property,
equivalent to the parallel axiom of this geometry is that the sum of the interior
angles of a triangle is always less than~$\pi$, the measure of the straight angle.
\vskip 7pt
\subsection{Tessellations in the hyperbolic plane}\label{tessellations}

   Henri Poincar\'e also established an important theorem from which we know
that there is an infinite family of tilings in the hyperbolic plane when in the 
Euclidean plane the same definition leads to three tilings only, up to similarity.

   To state the theorem, we have to remember the definition of a {\bf tessellation}.
Consider the following process. We start from a single polygon~$P$ and we replicate
it by reflection in its sides. Then, recursively, we replicate the images of~$P$ by
reflection in their sides. If the images do not overlap and if any point of the plane 
is contained in at least one image, we say that we have a tiling which is generated 
from~$P$ by tessellation. We also call the tiling a tessellation generated by~$P$.

   In the late 19$^{\rm th}$ century, Henri Poincar\'e proved the following:

\begin{thm}
A triangle~$T$ of the hyperbolic plane generates a tessellation if and only
if its angles are of the form
\hbox{$\displaystyle{{2\pi}\over p}$}, \hbox{$\displaystyle{{2\pi}\over q}$}
and \hbox{$\displaystyle{{2\pi}\over r}$}, where $p$, $q$ and~$r$
are positive integers satisfying
\hbox{$\displaystyle{1\over p}+\displaystyle{1\over q}
+\displaystyle{1\over r}< \displaystyle{1\over2}$}.
\end{thm} 

   Note that the condition merely says that $T$~is a triangle of the hyperbolic plane.

   An important particular case of this theorem is the case when $P$~is a regular
polygon. This case can be derived from the theorem by considering the rectangular
triangle~$T$ which is constructed from~$P$ by taking as vertices the centre of~$P$ 
and the end-points of a half-side. If $p$~is the number of sides of~$P$ and
if \hbox{$\displaystyle{{2\pi}\over q}$} is its interior angle, then the
angles of~$T$ are \hbox{$\displaystyle{\pi\over q}$}, \hbox{$\displaystyle{\pi\over p}$}
and \hbox{$\displaystyle{\pi\over 2}$}. Accordingly, the condition is
now \hbox{$\displaystyle{1\over p}+\displaystyle{1\over q} < \displaystyle{1\over2}$}.

   Remark that, in the Euclidean plane, we can perform the same construction
starting from a rectangular triangle~$T$. This time, as the sum of angles
of a triangle in the Euclidean plane is~$\pi$, we get that necessarily,
\hbox{$\displaystyle{1\over p}+\displaystyle{1\over q} = \displaystyle{1\over2}$},
which gives three solutions exactly: $p=q=4$, $p=6$ with $q=3$ and $p=3$
with $q=6$. This gives us the square, the regular hexagon and the equilateral triangle
respectively. 

   In the hyperbolic plane, we have an infinite family. Moreover, it is defined
with $p=3$ and then $q\geq 7$, or with $q=3$ and then $p\geq 7$ or when
$p\geq 4$ and then $q\geq 5$ or when $q\geq 4$ and then $p\geq 5$.
 
   Remember that in the case $\{5,4\}$ called the {\bf pentagrid}, very simple tools 
to navigate in this tiling were devised, see~\cite{mmJUCSii}. This was extended
to the tilings $\{p,4\}$ in~\cite{mmgsJUCSi}, then to the tiling $\{7,3\}$ called
the heptagrid in~\cite{ibkmACRI}, later to the tilings $\{p,q\}$
in~\cite{mmgsSCIpq}. All these tilings fall under the class of {\bf combinatoric tilings}
defined in~\cite{mmSCIcombi,mmBolyaiCluj}. In this class,
we have an algorithmic way to study the tiling which is based on the construction
of a tree which spans the tiling. The tree is associated to a process of
partition of the hyperbolic plane which generates several tiles at each step
of the construction. The construction produces the whole tiling in infinite time.
It is important to notice that there is a bijection between the tree and the tiles
of the tiling. A particular way to number the nodes of the tree provides us with
efficient tools of navigation in the tiling initially established in~\cite{mmJUCSii} 
for the pentagrid and then generalized as above indicated.

A precise description of these results and their proofs together
with the references can be found in~\cite{mmbook1}. What we remind in 
Section~\ref{whenevenq} will give an insight in the method and its results.

   In our next section, we turn to the general case of the tilings
$\{p,q\}$ when $q$~is even. Later, in Section~\ref{whenoddq}, we introduce the 
new idea and its application
to an appropriate splitting of the hyperbolic plane leading to another
construction of the tilings $\{p,q\}$ when $q$ is odd.

\section{The case $\{p,q\}$ when $q$ is even}
\label{whenevenq}

   Here, we remind the splitting of the hyperbolic plane thoroughly explained and analyzed
in~\cite{mmbook1}. It  leads to an algorithmic construction, in infinite time, of the 
tiling $\{p,q\}$ when $q$~is even. The case when $q$~is odd, which is more complex,
will be dealt with in the next section.

When $q$~is even, we define a {\bf sector}~${\cal S}_0$ as the angular 
sector defined by taking a vertex~$V$ of the polygon~$P$ on which the tiling is 
constructed and the rays issued from~$V$ which supports the two edges of~$P$ which 
meet at~$V$. We call $V$ the {\bf vertex} of~${\cal S}_0$ and $P$ is its {\bf head}. It is 
easy to see that the whole tiling is the union of $q$~copies of~${\cal S}_0$ which 
share the same vertex.

   Number the edges of~$P$ $e_1$,..., $e_p$, the numbers increasing while clockwise
turning around~$P$. Similarly, denote by~$V_i$,...,$V_p$ the vertices of~$P$, deciding
that $V_1$ and~$V_2$ are the end-points of~$e_1$. We also assume that the lines supporting
$e_1$ and~$e_p$ also support the rays~$\rho_\ell$ and~$\rho_r$ issued from~$V_1$
which define ${\cal S}_0$. As $q$~is even, we can write $q=2h$ and we can easily see 
that the angle between $e_2$ and~$\rho_r$ which is outside~$P$ and inside ${\cal S}_0$ is 
$(h$$-$$1)\displaystyle{{2\pi}\over q}$, see Figure~\ref{split_evS0}. 
And so, $h$$-$1 copies of ${\cal S}_0$ exactly fill up the region~$R_1$ which is 
outside~$P$, inside~${\cal S}_0$ and between~$\rho_\ell$ and the line which 
supports~$e_2$. Say that $R_1$ is a {\bf fan} of $h$$-$1 copies of~${\cal S}_0$.
We can define similar regions $R_i$ which are fans of $h$$-$1 copies of~${\cal S}_0$,
with $i\in\{2..p$$-$$3\}$. Indeed, $R_{i+1}$ is bordered by the continuation of~$e_i$
and the side~$e_{i+1}$ and its continuation outside~$P$ and outside~$R_i$.
Now, the complement in ${\cal S}_0$
of~$P$ and all the $R_i$'s we have just defined is a new region~$S_1$. This region is
defined by~$e_{p-2}$, $e_{p-1}$ and~$e_p$, the angles between $e_{p-1}$ and its
neighbouring edges inside~${\cal S}_1$ being both
$(h$$-$$1)\displaystyle{{2\pi}\over q}$.

\vskip 7pt
\setbox110=\hbox{\epsfig{file=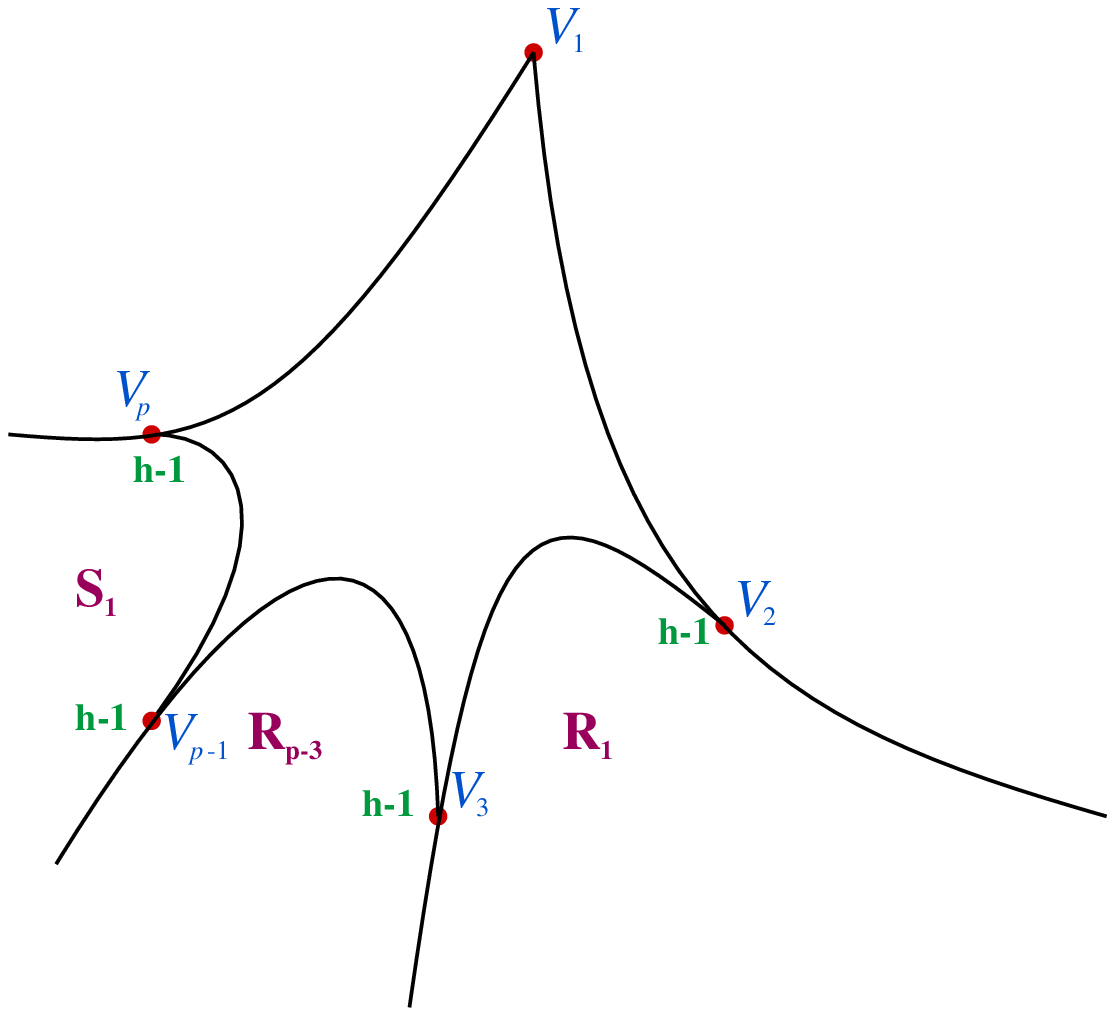,width=300pt}}
\vtop{
\ligne{\hfill
\PlacerEn {-150pt} {0pt} \box110
\hfill}
\vspace{-10pt}
\begin{fig}
\label{split_evS0}\small
Case when $q$~is even. The basic region ${\cal S}_0$ and its splitting.
\end{fig}
}

Let $P_1$ be the reflection of~$P$ in~$e_{p-1}$. Denote by~$\rho_\ell$ the continuation
of $e_{p-2}$ and by~$\rho_r$ that of~$e_p$. The splitting of~${\cal S}_1$ is
a bit different from that of~${\cal S}_0$, but it relies on the same considerations.
Rename~$e_1$ the edge of~$P_1$ which is shared with~$P$ and denote by~$e_i$
the other sides of~$P_1$, $i\in\{1..p\}$, the numbering being increasing while
clockwise turning around~$P_1$. We notice that this time, in the
complement in~${\cal S}_1$ of~$P_1$, $\rho_\ell$ and $e_2$ define an angle
which is $(h$$-$$2)\displaystyle{{2\pi}\over q}$, so that we split this
region, say~$R_1$ into~$(h$$-$$2)$ copies of~${\cal S}_0$: this time, we have a fan
of~$(h$$-$$2)$ copies of~${\cal S}_0$. The next regions $R_i$ are defined by~$e_{i}$ 
and~$e_{i+1}$ in the complement in~${\cal S}_1$ of~$P$ and the regions~$R_j$ for $j<i$,  
$e_{i-1}$ and~$e_i$. But, for $i\in\{2..p$$-$$3\}$, each $R_i$ is a fan of 
$h$$-$$1$ copies of~${\cal S}_0$. Now, the other ray~$\rho_r$ and~$e_p$
define in the complement of~$P$ in~${\cal S}_1$ another region~$R_{p-2}$ which
is also a fan of~$(h$$-$$2)$ copies of~${\cal S}_0$: this copies are obtained by
using an odd number of reflections in lines. Now, what remains
in~${\cal S}_1$ after removing~$R_{p-2}$ is a region~$R_{p-1}$ which is a copy 
of~${\cal S}_1$, see Figure~\ref{split_evS1}.
Accordingly, we have split ${\cal S}_1$ into a copy of~$P$, $(p$$-$$2)(h$$-$$1)-2$
copies of~${\cal S}_0$ and one copy of~${\cal S}_1$. 

   Note that we can write the splitting as follows:
\vskip 5pt
\ligne{\hfill
${\cal S}_0 \longrightarrow (p$$-$$3)(h$$-$$1).{\cal S}_0 + S_1$\hfill}
\ligne{\hfill
${\cal S}_1 \longrightarrow ((p$$-$$2)(h$$-$$1)$$-$$2).{\cal S}_0 + S_1$,\hfill}
\vskip 5pt
\noindent
which is the splitting given in~\cite{mmbook1}.

\setbox110=\hbox{\epsfig{file=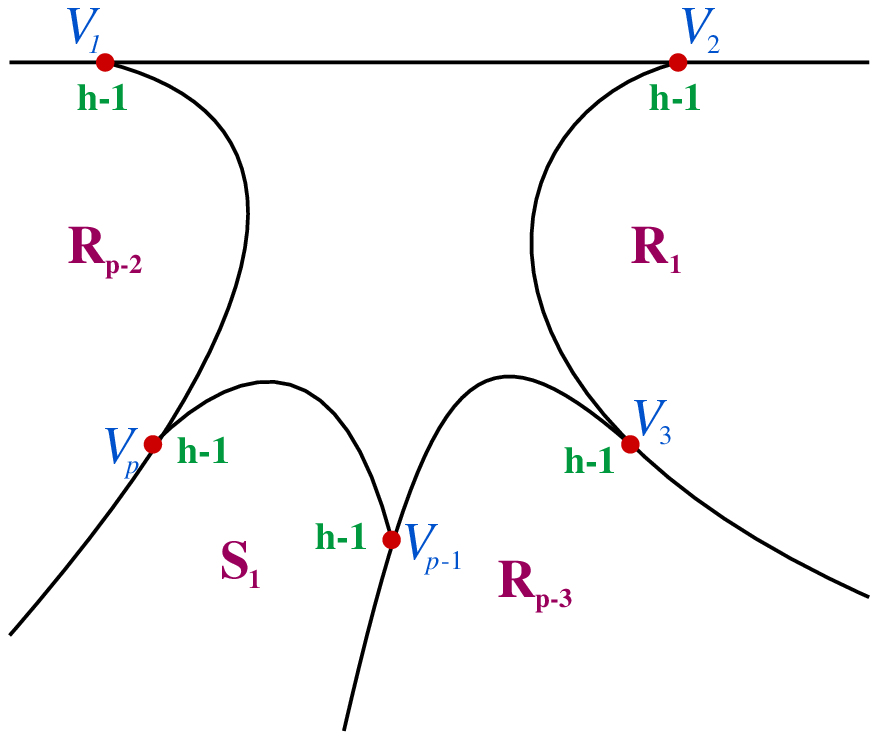,width=300pt}}
\vtop{
\vspace{-60pt}
\ligne{\hfill
\PlacerEn {-150pt} {0pt} \box110
\hfill}
\vspace{-45pt}
\begin{fig}
\label{split_evS1}\small
Case when $q$~is even. The basic region ${\cal S}_1$ and its splitting.
\end{fig}
}

   We briefly remind the consequence of the just computed relations. From them, 
we easily derive a matrix which we interpret as an {\bf incident matrix} called
the {\bf matrix of the splitting}. The rows indicate how a region is split in terms
of the basic regions, ${\cal S}_0$ and~${\cal S}_1$. The columns indicate how many
copies of the considered region enter the splitting of the region associated to the
considered line. From the matrix, we get its characteristic polynomial which here
we call the {\bf polynomial of the splitting}. It is not difficult to see that the
polynomial is \hbox{$P(X) = X^2-((p$$-$$3).(h$$-$$1)$$+$$1).X - h+3$}. It is known that,
this polynomial has a real root~$\beta$ which is positive and greater than~1, 
see~\cite{mmbook1}.
It is also known that the recurrent relation obtained from the polynomial defines
a sequence of increasing positive numbers in which we can decompose any natural number
with sums of terms of the sequence, each term entering the sum being multiplied by an
integer whose range is 0..$b$, where $b=\lfloor\beta\rfloor$. It is known that, in
general, the representation is not unique. However, it can be made unique by requiring
that we take the longest representation in terms of number of digits. We call
the set of these maximal representations the {\bf language of the splitting}. Now, it 
is also known that the language is regular if and only if $\beta$ is a Pisot number,
which means that $\beta>1$ and that if $\alpha$ is the other root, $\vert\alpha\vert<1$.
We know that most of the languages associated to a tiling~$\{p,q\}$ are regular
but they are very different. We refer the reader to~\cite{mmbook1} for the corresponding
study.

\section{The case $\{p,q\}$ when $q$ is odd}
\label{whenoddq}

   When $q$~is odd, we can no more split the tiling using lines which support the sides 
of the tiles. In~\cite{mmbook1}, we solved the problem by using a zig-zag line which follows
sides of the tiles. In Subsection~\ref{basic}, we remind the main lines
of this splitting and its main properties. In Subsection~\ref{split1}, we describe
the new idea and a way of splitting which we can straightforward devise from it.

\subsection{The previous solution}\label{basic}

   In~\cite{mmbook1} we define the basic regions by using a zig-zag line constituted
of sides of the tiling. Each side is defined from the previous one by making a constant 
angle defined by $h\displaystyle{{2\pi}\over q}$, where 
$h=\lfloor\displaystyle{q\over 2}\rfloor$.
This angle is the biggest positive integral multiple of
$\displaystyle{{2\pi}\over q}$ which is less than the straight angle. More
precisely, fix an edge~$e_1$. Let~$V$ be the vertex to which we arrive after choosing~$e_1$.
Denote the other vertices by $e_2$,~...,~$e_q$. The angle between $e_i$ and~$e_{i+1}$
is $\displaystyle{{2\pi}\over q}$ for $i\in \{1..q$$-$$1\}$ and it is the same for the
angle between $e_q$ and~$e_1$. Now, it is plain that
$e_{h+1}$ and $e_{h+2}$ are the closest to the line which supports~$e_1$: both these
edges make an angle $h\displaystyle{{2\pi}\over q}$ with~$e_1$. We decide to number
the edges in the following way: the last chosen edge is always~$e_1$. The other edges
are numbered from~2 up to~$q$ by counter clockwise turning around~$V$, the vertex
to which $e_1$~arrives and from which the edges $e_i$ are issued, $i\in\{2..q\}$. The next 
vertex is $e_{h+1}$: it becomes the new value of~$e_1$ and its other end becomes the new 
value of~$V$. See Figure~\ref{vertex_0} which illustrates this process.

\vskip 7pt
\setbox110=\hbox{\epsfig{file=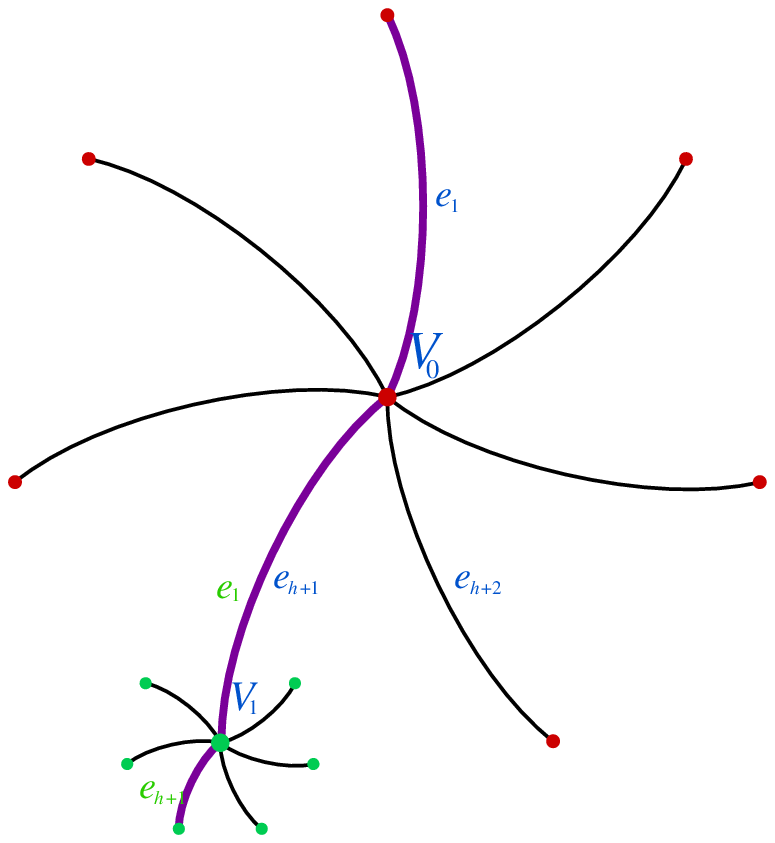,width=200pt}}
\vtop{
\ligne{\hfill
\PlacerEn {-100pt} {0pt} \box110
\hfill}
\vspace{-30pt}
\begin{fig}
\label{vertex_0}\small
Illustration of the choice of the next side in the zig-zag line devised
for the case when $q$~is odd. The resulting zig-zag line is in purple.
\end{fig}
}

   With these conventions, it is possible to define two regions ${\cal S}_0$
and ${\cal S}_1$ which provide results very close to those obtained in the case when
$q$~is even. Figures~\ref{split_oddS0} and~\ref{split_oddS1} illustrate the situation
in the case when $q=7$. This is enough to allow us to see how the splitting is performed in 
this case. In particular, we find the same splitting matrix as in the case
when $q$~is even.
 
\setbox110=\hbox{\epsfig{file=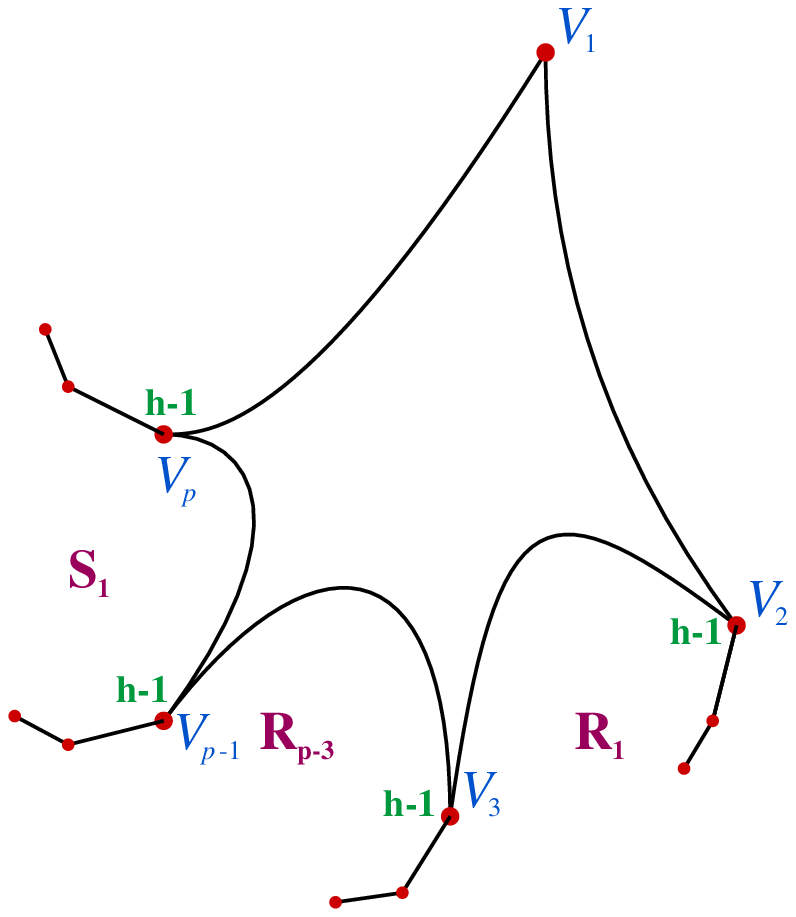,width=220pt}}
\vtop{
\vspace{-10pt}
\ligne{\hfill
\PlacerEn {-150pt} {0pt} \box110
\hfill}
\vspace{-30pt}
\begin{fig}
\label{split_oddS0}\small
Case when $q$~is odd. The basic region ${\cal S}_0$ and its splitting in the solution
given in~{\rm\cite{mmbook1}}.
\end{fig}
}

\setbox110=\hbox{\epsfig{file=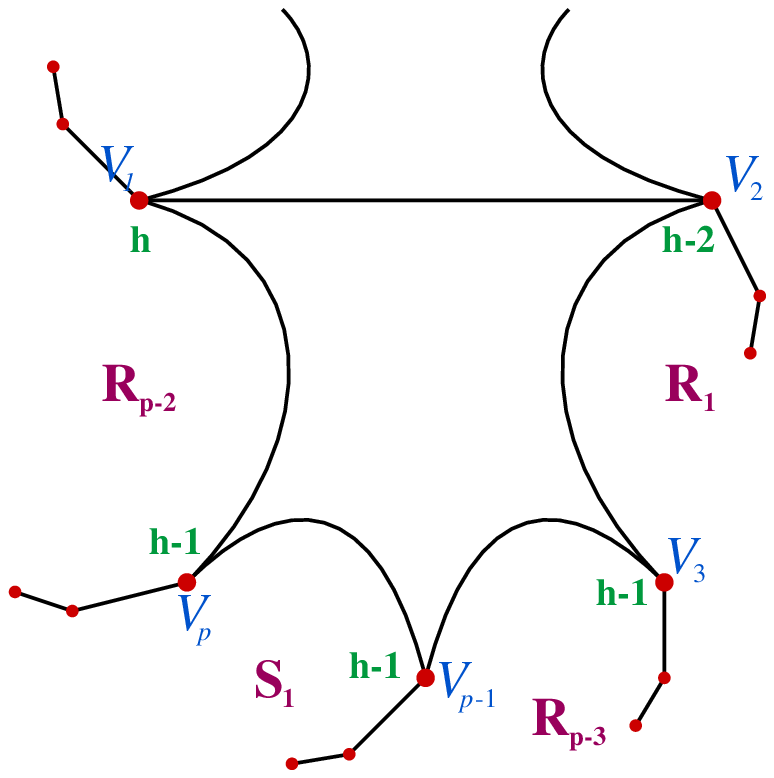,width=220pt}}
\vtop{
\vspace{-10pt}
\ligne{\hfill
\PlacerEn {-150pt} {0pt} \box110
\hfill}
\vspace{-50pt}
\begin{fig}
\label{split_oddS1}\small
Case when $q$~is odd. The basic region ${\cal S}_1$ and its splitting in the solution
given in~{\rm\cite{mmbook1}}.
\end{fig}
}

\subsection{The new splittings}\label{split1}

   Now, we turn to the new splittings announced in the abstract and in 
Section~\ref{intro}. Both the new splittings are based on the same notion
of a mid-point line.
   
   This notion was introduced in the heptagrid, in order to
define the regions of the splitting giving rise to this tiling. In this case,
a mid-point line is a line which joins mid-points of consecutive edges of the tiling.

\vskip 7pt
\setbox110=\hbox{\epsfig{file=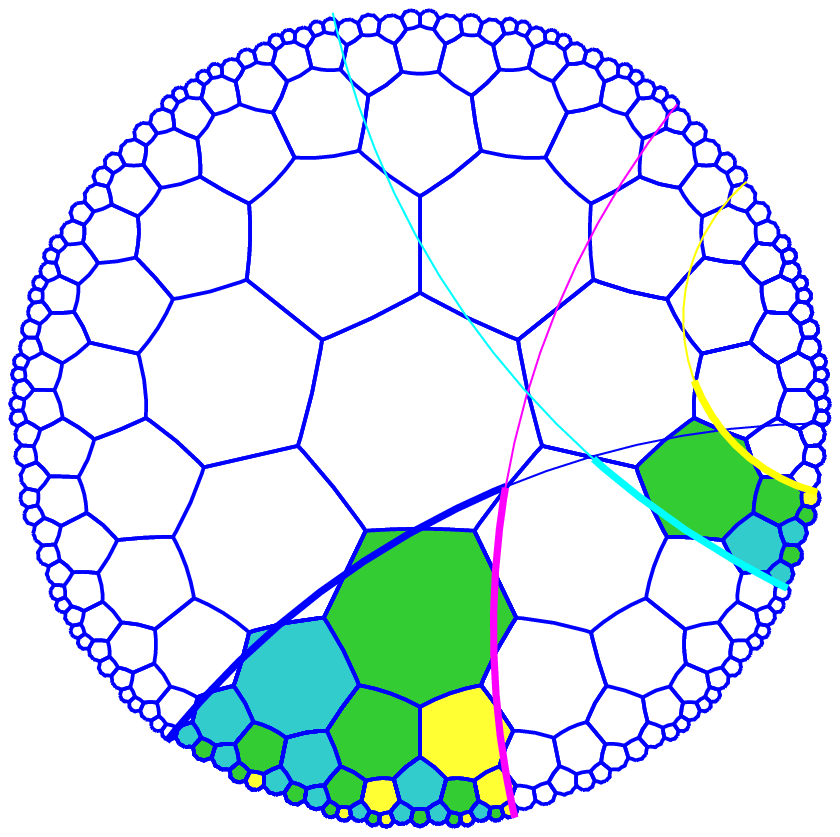,width=220pt}}
\vtop{
%\vspace{-10pt}
\ligne{\hfill
\PlacerEn {-100pt} {0pt} \box110
\hfill}
%\vspace{-10pt}
\begin{fig}
\label{mid_point_lines}\small
The mid-point lines of the heptagrid. They allow us to define the two basic regions
giving rise to this tiling.
\end{fig}
}

   It is not difficult to see that when $q\geq5$, this property of the mid-points of 
edges is no more true: three mid-points $M_1$, $M_2$ and $M_3$ such that $M_1$ with~$M_2$
and $M_2$ with $M_3$belong to the same heptagon but $M_1$ and $M_3$ belong to different
ones are never on the same line. In the next subsubsection, we shall see that however, 
there is a way to generalize the mid-points of the heptagrid. Then, in 
Subsubsection~\ref{splitnew1} and in Subsection~\ref{split2} we shall see the application
of this new construction to two variants of a splitting of the tiling $\{p,q\}$ in
the case when $q$~is odd. Subsubsection~\ref{til45} will deal in another way for the special
case when $p=4$ and $q=5$.

\subsubsection{The mid-point lines}
   Let $h=\lfloor\displaystyle{q\over 2}\rfloor$, as this number will play an important 
role. In our new setting, we shall again use a mid-point line, but 
this time, it will be defined by the angle $h\displaystyle{{2\pi}\over q}$. Note that
when $q=3$, $h=1$, so that this definition is a natural generalization of what we did
in the heptagrid. Indeed, consider~$V$ a vertex of~$P$, a regular polygon with 
$p$~sides and $q$~copies of~$P$ exactly covering a neighbourhood of a vertex.
Let $Q_i$ be the copies around~$V$ covering one of its neighbourhoods. Assume that
$Q_1=P$ and let the others, $Q_2,...,Q_p$ be increasingly numbered while counter clockwise
turning around~$V$. Let~$e_0$ be the side of~$P$ abutting~$V$ which is not shared by~$Q_2$.
The continuation of~$e$ cuts $Q_{h+1}$ into two parts which are the reflection of each other
in the continuation of~$e$. Let~$a$ be the side of~$Q_h$ which makes the 
angle~$h\displaystyle{{2\pi}\over q}$ with~$e$ at~$V$. Let $M$~be the mid-point of~$e$
and let $N$~be that of~$a$. Let $\delta$ be the line joining~$M$ and~$N$. Let~$W$ be the
other end-point of~$a$. Consider the angle at~$W$ defined by~$a$ which is
$h\displaystyle{{2\pi}\over q}$, but defined clockwise. This defines the side~$b$
of one copy of~$P$ among those which, dispatched around~$W$ exactly cover a neighbourhood
of this point. Now, from the above definition of angles, it is not difficult
to see that with respect to the line~$d$ supporting~$a$ we have that $e$ lies in one of the
half-plane delimited by~$d$ and $b$~lies in the other.
Let $O$ be the mid-point of~$b$. The triangles $MVN$ and $NWO$ are equal
and this shows that $\delta$ also joins~$N$ to~$O$, see Figure~\ref{vertex_new}
as $V$ and~$W$ are not on the same side of~$d$. In this way, we can see that
$\delta$~is a line of mid-points of sides of the tiles, but instead of joining mid-points
of consecutive edges, it joins mid-points of edges of the tiles which make the angle
$h\displaystyle{{2\pi}\over q}$: we shall say that the line joins {\bf $h$-consecutive
mid-points}. From now on, we shall call {\bf $h$-mid-point lines} the lines which joins
$h$-consecutive mid-points.

%\vskip 7pt
\setbox110=\hbox{\epsfig{file=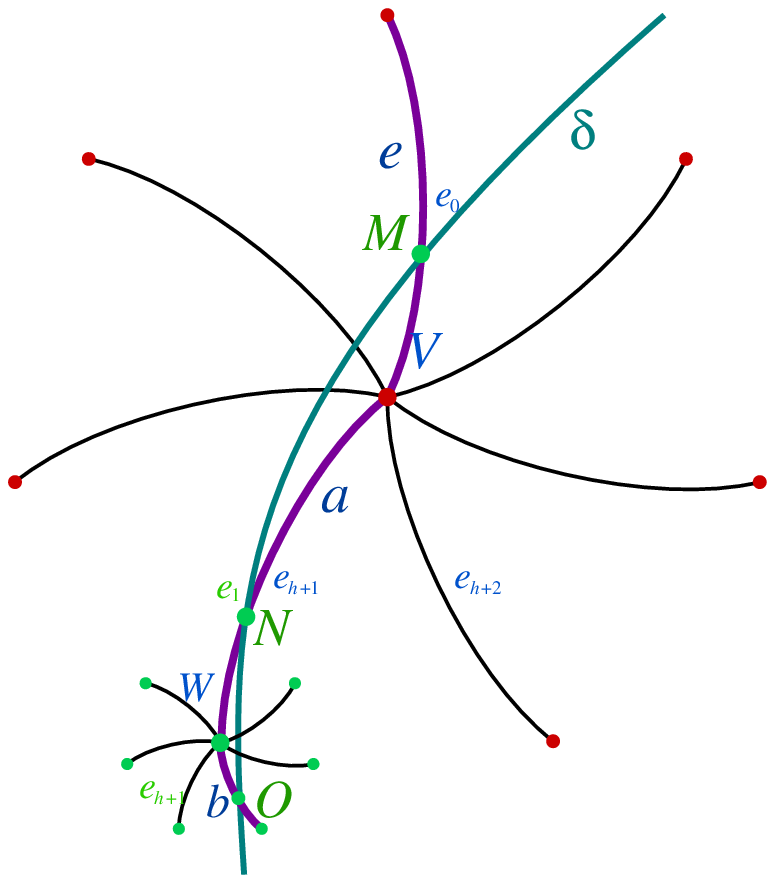,width=200pt}}
\vtop{
\ligne{\hfill
\PlacerEn {-100pt} {0pt} \box110
\hfill}
\vspace{-30pt}
\begin{fig}
\label{vertex_new}\small
Illustration of the construction of a $h$-mid-point line. Here, $q=7$ and so, $h=3$.
\end{fig}
}

\subsubsection{The splitting}\label{splitnew1}

   Now, we are ready to define the splitting.

   First, define a sector~${\cal S}_0$ to be delimited by a vertex~$V$, called the 
vertex of~${\cal S}_0$. The head of the sector is a copy of~$P$ for which~$V$ is a vertex.
Let $b$ and~$c$ denote the sides of~$P$ which meet at~$V$. The bisector of the angle
between $b$ and~$c$ at~$V$ which is outside~$P$ is the edge~$a$ of a copy of~$P$ among
the $q$~copies of~$P$ which can be put around~$V$. Let $C$, $B$ and~$A$ be the mid-points 
of~$c$, $b$ and~$a$ respectively. We may assume that we go from~$c$ to~$b$ by counter 
clockwise turning around~$V$. Let $\rho_B$ be the ray issued from~$B$ whose supporting
line goes through~$A$. We take the ray which does not contain~$A$. Note that the ray is
supported by a $h$-mid-point line. Now, we define~$\rho_C$ by taking the image 
of~$\rho_B$ under the rotation around~$V$ which transforms~$B$ into~$C$. Clearly,
$\rho_C$ is also a $h$-mid-point line: the next mid-point on this ray, starting from~$C$
belongs to an edge of a tile which is outside~$P$, see Figure~\ref{sect_S0}. 
We define ${\cal S}_0$ as the region delimited by $V$, $B$, $C$ and both rays $\rho_B$ 
and~$\rho_C$. From the definition of~$\rho_B$, we easily conclude that 
the tiling $\{p,q\}$ can exactly be split into $q$~copies of~${\cal S}_0$. 

\setbox110=\hbox{\epsfig{file=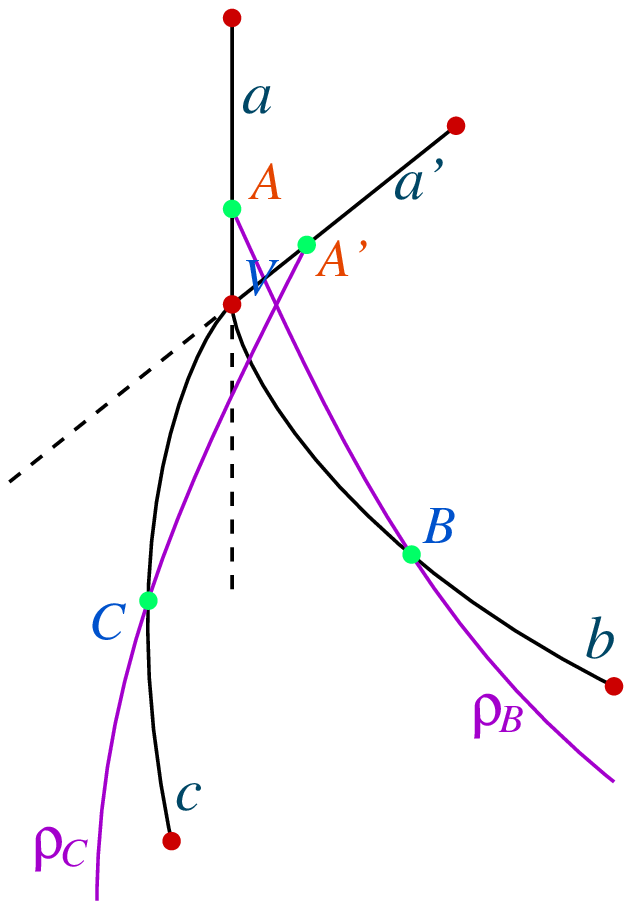,width=200pt}}
\vtop{
\vspace{-30pt}
\ligne{\hfill
\PlacerEn {-280pt} {0pt} \box110
}
\vspace{-10pt}
\begin{fig}\label{sect_S0}
\leurre
The $h$-mid-point rays used for the definition of the region ${\cal S}_0$.
\end{fig}
}

   Consider the region ${\cal S}_0$, as illustrated in Figure~\ref{split_S0}. The figure 
indicates how we split this region. There is a small difference with the case when $q$~is
even: we have to introduce three regions in order to get a combinatoric splitting.

   First, consider the delimitation of the sector. On the figure, the mid-points
are called $M_1$, $M_2$, ..., $M_p$ as most of them are used in the splitting. Here, $M_2$
plays the role of~$C$ and $M_1$ that of~$B$. On the figure, for each vertex $V_i$, we have 
represented the side~$s_i$ of the polygon having~$V_i$ among its vertices which makes the
angle $h\displaystyle{{2\pi}\over q}$ with the edge~$e_{i-1}$ of~$P$, with $i\in\{2..p\}$
and we consider~$e_p$ when $i=1$. We consider the ray issued from~$M_i$ and which goes
through the mid-point~$m_i$ of~$s_i$. On Figure~\ref{split_S0}, we can see that $V_2$ is
outside~${\cal S}_0$. However, the $h$$-$1 copies of~$P$ which share~$V_2$ and which
are inside the angle between $e_1$ and~$s_2$ are considered as belonging to~${\cal S}_0$:
for these polygons, $V_2$ is the single vertex among those of the polygon which is not
in~${\cal S}_0$. This is why we define the set of tiles spanned by~${\cal S}_0$ as those
which have at most one vertex outside~${\cal S}_0$.
 
\setbox110=\hbox{\epsfig{file=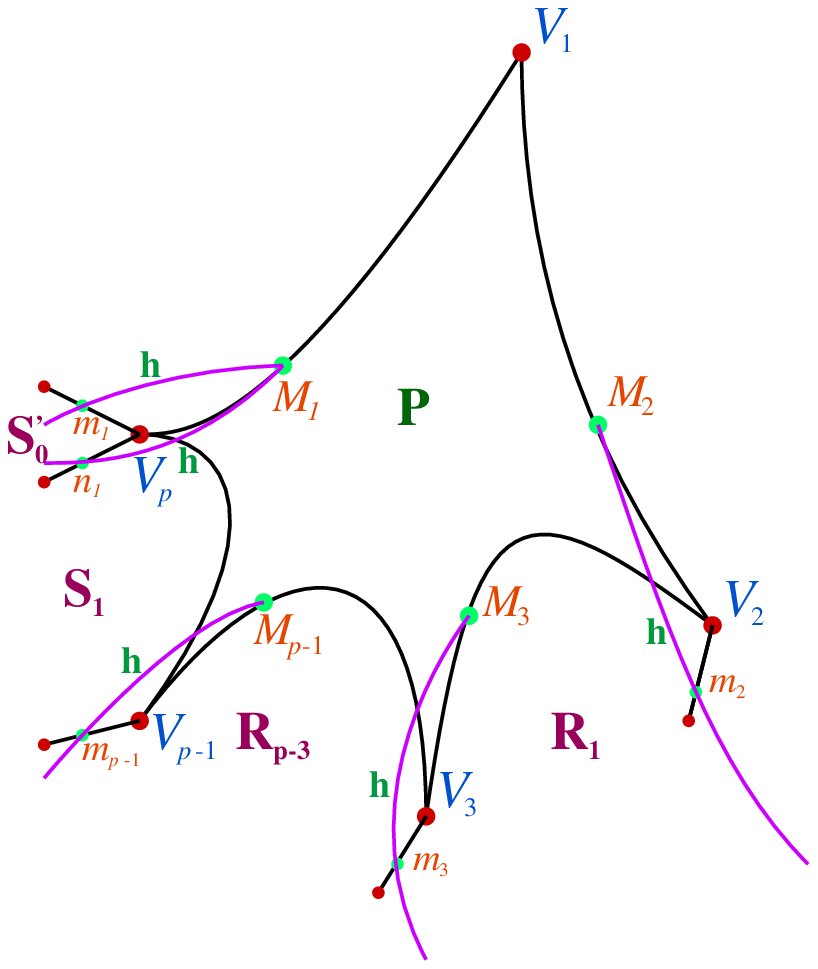,width=300pt}}
\vtop{
\ligne{\hfill
\PlacerEn {-330pt} {0pt} \box110
}
\vspace{-35pt}
\begin{fig}\label{split_S0}
\leurre
Splitting the region $S_0$.
\end{fig}
}

   Now, from this remark, we can say that there is a fan of $h$$-$1 copies of ${\cal S}_0$
delimited by $V_2$, the ray issue from~$m_2$ and supported by that joining~$M_2$
to~$m_2$ and the ray issued from~$M_3$ and passing through~$m_3$. The just described 
region~$R_1$ consists of $h$$-$1 copies of ${\cal S}_0$, $V_2$ being their common vertex.
Similar regions $R_2$, ..., $R_{p-3}$ can successively be defined in the complement
in ${\cal S}_0$ of~$P$ and the regions already defined.

   When we arrive to~$M_{p-1}$, the ray issued from~$M_{p-1}$ passing 
through~$m_{p-1}$ defines the left-hand side of~$R_{p-3}$. What remains from~${\cal S}_0$
is a region which we split as indicated in Figure~\ref{split_S0}. From~$M_1$ we draw the
two rays $\rho_m$ and~$\rho_n$ which are supported by the $h$-mid-point lines passing 
through~$M_1$. One ray goes through~$m_1$ and the other from~$n_1$. Now, the region 
delimited by~$M_1$ and the rays~$\rho_m$ and~$\rho_n$ is a new type of region which we
call ${\cal S}'_0$ as it looks like~${\cal S}_0$. Now, what remains from~${\cal S}_0$
once we removed ${\cal S}'_0$ is by definition~~${\cal S}_1$.

   And so, we can summarize the splitting of~${\cal S}_0$ by writing:
\vskip 5pt
\ligne{\hfill
${\cal S}_0 \longrightarrow (p$$-$$3)(h$$-$$1){\cal S}_0 + {\cal S}'_0 + {\cal S}_1$.
\hfill}

   Now, as ${\cal S}'_0$ is obtained from~${\cal S}_0$ by just removing a copy 
of~${\cal S}'_0$, we immediately get that the splitting of~${\cal S}'_0$, illustrated by
Figure~\ref{split_Sb0} can be summarized by the following formula: 

\vskip 5pt
\ligne{\hfill
${\cal S}'_0 \longrightarrow (p$$-$$3)(h$$-$$1){\cal S}_0 + {\cal S}_1$.
\hfill}

\setbox110=\hbox{\epsfig{file=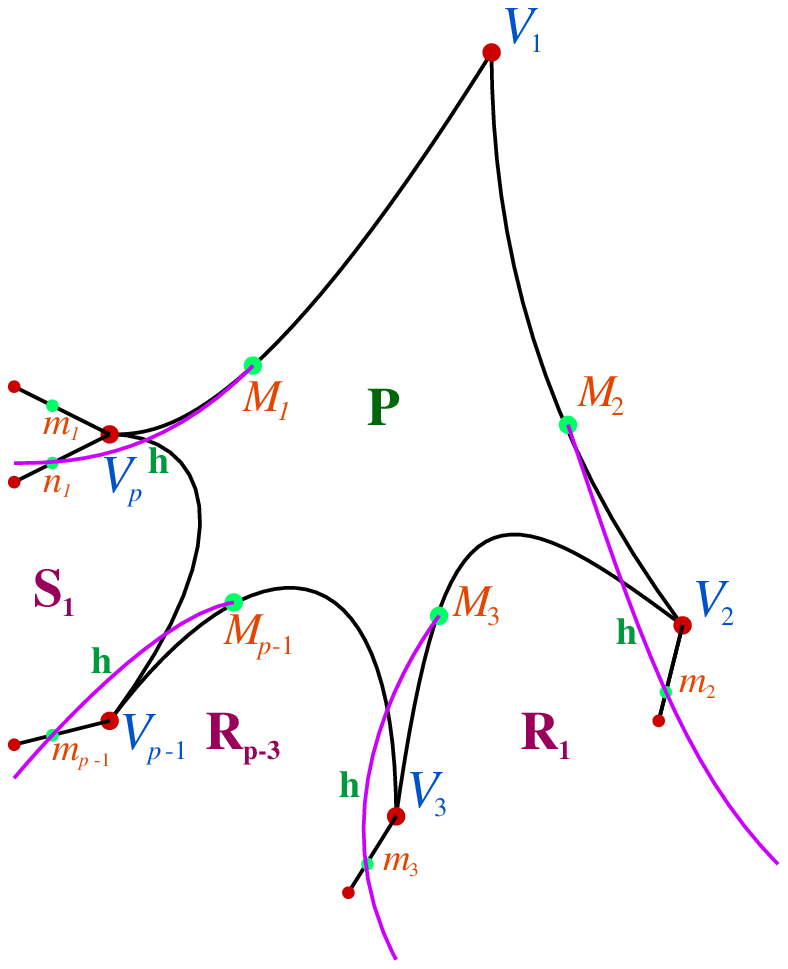,width=300pt}}
\vtop{
\ligne{\hfill
\PlacerEn {-330pt} {0pt} \box110
}
\vspace{-35pt}
\begin{fig}\label{split_Sb0}
\leurre
Splitting the region $S'_0$.
\end{fig}
}

   Presently, we arrive at the splitting of~${\cal S}_1$. Everything goes as 
for~${\cal S}_0$, taking into account the following: for defining~${\cal S}_1$,
we need $p$$-$1 sides of~$P$, but we have to take into account the mid-points of the edges
of the reflection of~$P$ in~$V_1V_2$ which have $V_1$ and~$V_2$ as end-points, see 
Figure~\ref{split_S1}. We start the splitting both from~$V_2$ and from~$V_1$.
From~$V_2$, we define a region~$R_1$ as previously. It is a fan of $h$$-$2 copies 
of~${\cal S}_0$ instead of $h$$-$1 because ${\cal S}_1$ is defined from~$M_2$ which,
in this case, is not on an edge of~$P$ but on an edge of the reflection of~$P$ in~$V_1V_2$.
The $h$$-$2 polygons having~$V_2$ as a vertex are counted as included in~${\cal S}_1$
for the same reason as we did in the case of~${\cal S}_0$. Now, this remark also holds
for~$V_1$: from there we define a fan of $h$$-$2 copies of~${\cal S}_0$, defining a 
region $R_ {p-2}$, whose right-hand side limit is defined by the ray issued from~$M_p$
and passing through~$m_p$. Note that the copies of ${\cal S}_0$ which constitute
$R_{p-2}$ are obtained from those which constitute $R_1$ by a reflection in the
bisector of~$V1V2$. 

\vskip 5pt
\setbox110=\hbox{\epsfig{file=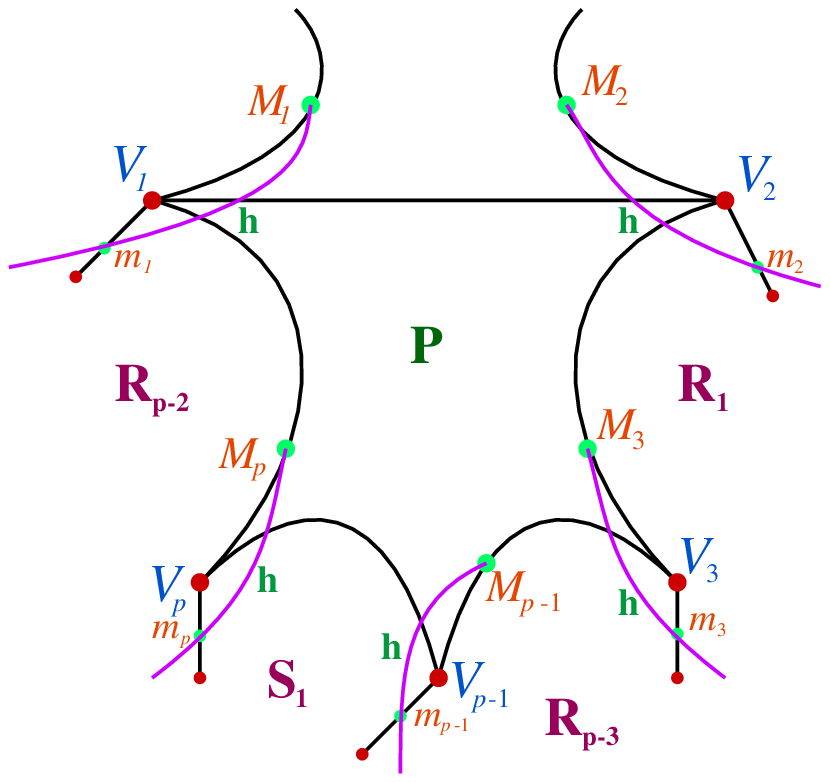,width=300pt}}
\vtop{
\vspace{-25pt}
\ligne{\hfill
\PlacerEn {-330pt} {0pt} \box110
}
\vspace{-85pt}
\begin{fig}\label{split_S1}
\leurre
Splitting the region $S_1$.
\end{fig}
}
\vskip 5pt
   As we did in the case of~${\cal S}_0$, we can define the regions $R_2$, ..., $R_{p-3}$
which are fans of $h$$-$1 copies of~${\cal S}_0$. Now, when we arrive at~$R_{p-3}$,
what we have between this latter region, $P$ and~$R_{p-2}$ is a copy of~~${\cal S}_1$.
Accordingly, we can summarize the splitting by:
\vskip 5pt
\ligne{\hfill
${\cal S}_1 \longrightarrow ((p$$-$$2)(h$$-$$1)-2){\cal S}_0 + {\cal S}_1$.
\hfill}
\vskip 5pt
\noindent
Indeed, in this formula, the term $(p$$-$$2)(h$$-$$1)-2$ comes from $p$$-$4 fans
with $h$$-$1 copies of~${\cal S}_0$, and the two fans with $h$$-$2 copies of~${\cal S}_0$
defined by $R_{p-2}$ and~$R_1$. 

   Accordingly, the polynomial of the splitting is:
\vskip 5pt
\ligne{\hfill
$P(X) = X^3 - ((p$$-$$3)(h$$-$$1)$+$1)X^2 - ((p$$-$$2)(h$$-$$1)$$-$$2)X - h$+3.\hfill}

   From this, we know that the number~$u_n$ of nodes which are on the same level of the
spanning tree of the tiling is defined by:

\vskip 5pt
\ligne{\hfill
$u_{n+3} = ((p$$-$$3)(h$$-$$1)$+$1)u_{n+2} + ((p$$-$$2)(h$$-$$1)$$-$$2)u_{n+1}
- (h$$-$$3)u_n$.\hfill}
\vskip 5pt  
   Note that easy computations give us that $P($$-$$1) =\ $$-$2, $P(0) =\ $$-$$h$+3 and
$P($$-$$\displaystyle{1\over2}) = \displaystyle{\hbox{$(p$$-$$5)$}\over4}(h$$-$$1)%
+\displaystyle{5\over8}$. Accordingly, when $h>3$ and $p\geq5$, we have that 
$P($$-$$\displaystyle{1\over2}) > 0$ and $P(0) < 0$. This shows that $P$~has three
real roots, that two of them are in $]$$-$$1,1[$. Let $a=(p$$-$$3)(h$$-$$1)+1$.
Then, an easy computation shows that $P(a) < 0$. This tells us that $\beta >a$
where $\beta$ is the greatest real root of~$P$. Now, under the assumption that
$h>1$, $a>1$, so that $\beta>1$ too. Accordingly, when $h\geq 4$ and $p\geq5$, $P$ is
a Pisot polynomial which means that the language of the splitting is regular.
%where the language of the splitting is defined by the representations of positive 
%numbers in the basis defined by the above sequence $\{u_n\}$ and which have the maximal
%number of digits in this basis. 
   
   We remain with the study of the cases when $h=2$ and $h=3$ one one hand and
the case when $p=4$ on the other hand.

   First, assume that $p\geq 5$.

   When $h=3$, $P(X)$ can be divided by~$X$ and so we can replace~$P$ by a
polynomial of degree~2 which we again call~$P$. We have:

\ligne{\hfill
$P(X) = X^2 -(2p$$-$$5)X - (2p$$-$$6)$.\hfill} 

   Note that the splitting assumes that we have $p\geq 4$. 
   In this case, we have two real roots, $\alpha$ and~$\beta$, assuming $\alpha<\beta$.
Now, $P((2p$$-$$5) =\ $$-$$2p$$+$$6 < 0$, so that $\beta> 2p$$-$$5 > 1$. On another hand, 
$P(0) = \ $$-$$2p$$+$$6 <0$ too. Now, $P($$-$$1) = 1 + 2p$$-$$5 - (2p$$-$$6) = 2$,
which means that $-$$1 < \alpha < 0$. Accordingly, $P$~is a Pisot polynomial
and the language of the splitting is again regular. 

   When $h=2$, the polynomial is now:

\ligne{\hfill
$P(X) = X^3 -(p$$-$$2)X^2 - (p$$-$$4)X + 1$.\hfill} 

   This time we have that $P(0) = 1 >0$. Remember that
$P($$-$$1)=\ $$-$$2 < 0$ as the computation does not depend neither on~$h$ nor on~$p$.
An easy computation gives us $P(1) = 8 - 2p < 0$ when $p\geq5$. Now,
$P(p$$-$$2) =\ $$-$$(p$$-$$4)(p$$-$$2) +1 < 0$ when $p\geq 5$, so that $\beta>p$$-$$2>1$,
where $\beta$ is the greatest real root of~$P$: again, $P$~is a Pisot polynomial.
 
This allows us to conclude that in all cases when $p\geq5$, we have that $P$~has
three real roots, that two of them have a modulus which is less than~1 and that the biggest
real root is positive and greater than~1. We have that $P$~is a Pisot polynomial and,
consequently, the language of the splitting is always regular in this case.

   Let us look at the case when $p=4$.

   This time we have that $P(X)= X^3 - hX^2 - 2(h$$-$$2)X-h+3$.

   Denote by $\beta$, $x_1$ and~$x_2$ the roots of~$P$, $\beta$ being the
greatest real root, which is positive. Indeed: $P(h) = $$-$$2(h$$-$$2)h-h+3=
$$-$$2h^2 + 3h +3 <0$ when $h\geq 3$, which implies that $h<\beta$. Now,
$\beta x_1x_2=\ $$-$$h$+3 and, on another hand, $x_2=\overline{x_1}$ as the coefficients
of~$P$ are real numbers. This gives the following computation: 
$\vert x_1\vert^2 = \displaystyle{\hbox{$h$$-$3}\over\beta}
< \displaystyle{\hbox{$h$$-$3}\over h} = 1 - \displaystyle{3\over h}$. Accordingly,
$\vert x_1\vert < 1$ so that $P$~is a Pisot polynomial when $h\geq3$.

   We remain with the single case $h=2$ under the assumption that $p=4$.

   In this case $P(X) = X^3 -2X^2 +1$. Clearly, 1 is a root of this polynomial. The other
root are those of $X^2 -X -1$ as $P(X)= (X$$-$$1)(X^2$$-$$X$$-$$1)$. The other roots of~$P$
are $\alpha$ and~$\beta$ with $\beta>1$ and $\alpha<1$, as 
$\beta=\displaystyle{{1+\sqrt5}\over2}$ and, consequently, 
$\alpha=\displaystyle{{\sqrt5-1}\over2}$. But, as 1 is a root of~$P$, $P$ is no more
a Pisot polynomial. As the language of the splitting is regular if and only
if its polynomial is a Pisot polynomial or the product of a Pisot polynomial with
polynomials of the form $\displaystyle{\sum\limits_{k=0}^m X^k}$, we have that
the language of the splitting is not regular when $p=4$ and~$h=2$.

   Accordingly, in all cases when the splitting holds, its language is regular,
except in the case when $p=4$ and $h=2$.

\subsection{A variant of the new splitting}\label{split2}

   A variant of the previous splitting consists in replacing the region ${\cal S}_0$
by the region ${\cal S}'_0$ and by keeping the region~${\cal S}_1$ unchanged.
We have two basic regions instead of three, but the coefficients of the splitting 
matrix are different.

   In order to see this point, let us again have a look at the situation around a vertex.
Figure~\ref{sect_Sb0} represents the rays which are used to delimit the region
${\cal S}'_0$ in a way which can be seen as a local zoom on the part of 
Figure~\ref{split_Sb0}.  

   Now, Figure~\ref{deuxsect_Sb0} shows two sectors ${\cal S}'_0$ headed by adjacent
polygons around a common vertex. In the figure, the right-hand side sector is delimited
by the $h$-mid-point lines $\rho_{AB}$ and~$\rho_{AC}$ while the left-hand side one
is delimited by the $h$-mid-point lines $\rho_{GC}$ and~$\rho_{GF}$. It can be seen that
the continuation of the rays $\rho_{AC}$ and~$\rho_{GC}$ issued from~$C$ define
again a copy of~${\cal S}'_0$ which contains exactly the tiles contained in this
angular sector and which are not contained neither in the sector defined by
$\rho_{AB}$ and~$\rho_{AC}$ nor that which is defined by $\rho_{GC}$ and~$\rho_{GF}$.

\setbox110=\hbox{\epsfig{file=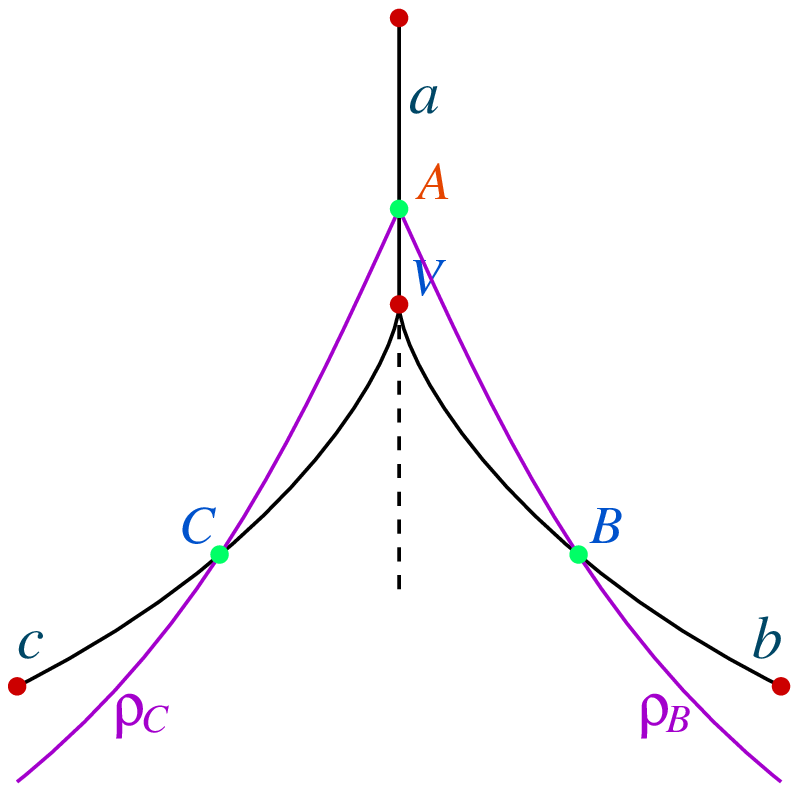,width=200pt}}
\vtop{
\vspace{-30pt}
\ligne{\hfill
\PlacerEn {-280pt} {0pt} \box110
}
\vspace{-25pt}
\begin{fig}\label{sect_Sb0}
\leurre
The $h$-mid-point rays used for the definition of the region ${\cal S}'_0$.
\end{fig}
}

   This means that, around a vertex, the $q$~polygons which meet at this vertex define
$q$~copies of~${\cal S}'_0$ which do not cover the hyperbolic plane. In between two
consecutive copies of~${\cal S}'_0$, there is room for exactly one copy. Accordingly,
we can split the hyperbolic plane into $2q$~copies of~${\cal S}'_0$, reproducing $q$~times
the scheme represented by Figure~\ref{deuxsect_Sb0}.

   Now, consider Figure~\ref{deuxsect_Sb0}. As the side $g$~is a rotated image of
the side~$a$, we get that $\rho_{GC}$ is a rotated image of~$\rho_{AB}$ under the same 
rotation. Now, we have just seen that the continuation of the rays $\rho_{AC}$ 
and~$\rho_{GC}$ issued from~$C$ define again a copy of~${\cal S}'_0$. This means
that ${\cal S}_0$ can be split into two copies of~${\cal S}'_0$ exactly. Accordingly,
the splittings of~${\cal S}'_0$ and ${\cal S}_1$ illustrated by Figures~\ref{split_Sb0}
and~\ref{split_S1} can be adapted to the regions ${\cal S}'_0$ and~${\cal S}_1$ only.

   This gives us the following formulas:

\vskip 5pt
\ligne{\hfill
${\cal S}'_0 \longrightarrow 2(p$$-$$3)(h$$-$$1){\cal S}'_0 + {\cal S}_1$.
\hfill}
\vskip 5pt
\ligne{\hfill
${\cal S}_1 \longrightarrow 2((p$$-$$2)(h$$-$$1)-2){\cal S}'_0 + {\cal S}_1$.
\hfill}
\vskip 5pt

\setbox110=\hbox{\epsfig{file=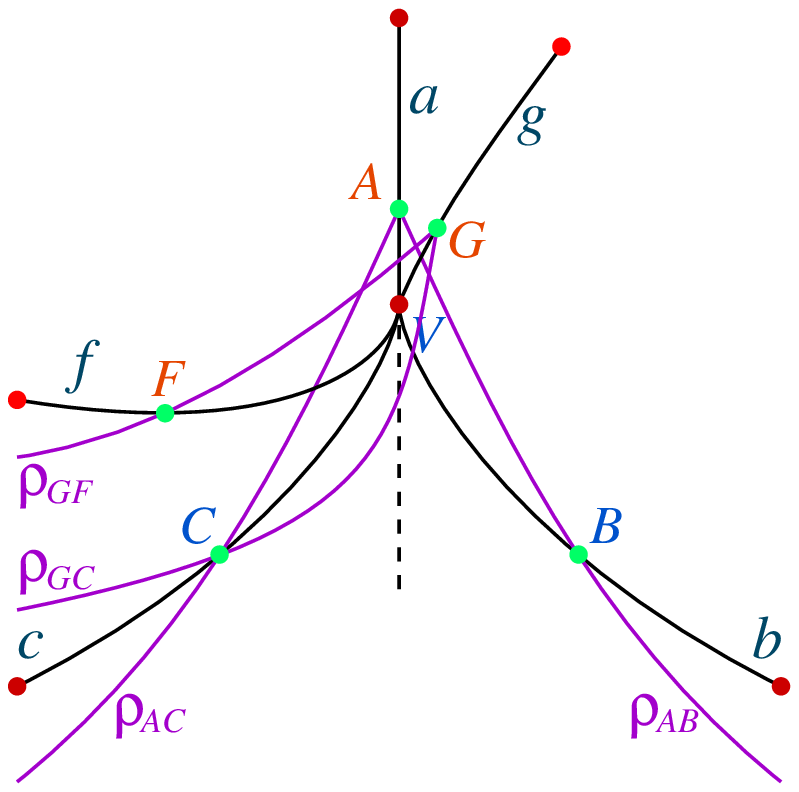,width=200pt}}
\vtop{
\vspace{-30pt}
\ligne{\hfill
\PlacerEn {-280pt} {0pt} \box110
}
\vspace{-25pt}
\begin{fig}\label{deuxsect_Sb0}
\leurre
Two consecutive copies of the region ${\cal S}'_0$.
\end{fig}
}

   This time, the polynomial of the splitting is:
\vskip 5pt
\ligne{\hfill
$P(X) = X^2 - (2(p$$-$$3)(h$$-$$1)$+$1)X - 2h$+6.\hfill}
\vskip 5pt
\noindent
As $q=2h$+1, this can be rewritten as
\vskip 5pt
\ligne{\hfill
$P(X) = X^2 - ((p$$-$$3)(q$$-$$3)$+$1)X - q$+7.\hfill}
\vskip 5pt

   When $q\geq 7$, the last coefficient of this polynomial of degree~2 is negative and
so it has a positive real root~$\beta$. We remark that 

$P((p$$-$$3)(q$$-$$3)) = -(p$$-$$3)(q$$-$$3) -q+7<0$, 

\noindent
as $q>7$ and $(p$$-$$3)(q$$-$$3) >0$ when $p\geq 4$ and $q> 7$.
so that $\beta> (p$$-$$3)(q$$-$$3)$. Consequently, if $\alpha$ is the other root
of the polynomial, it is also a real number and we have that

$\vert\alpha\vert\beta= q$$-$$7<q$$-$3, and so 
$\vert\alpha\vert<\displaystyle{\hbox{$q$$-$3}\over\hbox{($p$$-$$3)(q$$-$3)}}
=\displaystyle{1\over\hbox{$p$$-$3}}\leq1$,

\noindent
as $p\geq4$. And so, $\vert\alpha\vert<1$ which means that $P$~is a Pisot polynomial.

   We remain with the cases when $h=2$ and $h=3$.

   When $h=2$, the polynomial is now:
\vskip 5pt
\ligne{\hfill
$P(X)=X^2-(2(p$$-$$3)+1)X+2$.\hfill}
\vskip 5pt
 $P(2(p$$-$$3))=-2(p$$-$$3)+2 <0$ when $p\geq 5$. Accordingly, $P$~has a real root
$\beta$, with $\beta>2(p$$-$$3) >1$ when $p\geq 5$. There is another root $\alpha$
for which we have $\alpha\beta=2$, so that 
$\alpha=\displaystyle{2\over\beta} <\displaystyle{2\over{2(p-3)}}
=\displaystyle{1\over{p-3}} <1$ as $p\geq 5$. 

   When $p=4$ and $h=2$, we have that $P(X)=X^2 - 3X+2 = (X$$-$$1)(X$$-$$2)$.
As 1 is a root of~$P$, it cannot be a Pisot polynomial and so, the language of the
splitting is not regular in this case, as it is also the case with the splitting
of Subsection~\ref{split1}.

  When $h=3$, the polynomial becomes $P(X) = X^2-(2(p$$-$$3)$+$1)X$, so that the
roots are now $X=0$ and $X=2(p$$-$$3)$+$1>1$ when $p\geq4$. And so, in this case,
$P$ is a Pisot polynomial.

Accordingly, the language of the splitting defined by this new variant is regular
in all cases, except in the same case as that induced by the splitting of 
Subsection~\ref{split1}, {\it i.e.} when $p=4$ and $h=2$. 

\subsubsection{The special case of the tiling $\{4,5\}$}
\label{til45}

For this special case, corresponding to the case when $p=4$ and $h=2$ in our study of
Subsections~\ref{split1} and~\ref{split2}, it can be noted that the tiling $\{4,5\}$ 
is the dual tiling of the tiling $\{5,4\}$, the pentagrid for which 
the language of the splitting is 
regular. From this remark, we deduce that the coordinates which are used for the 
pentagrid can also be used for the tiling~$\{4,5\}$. In order to show how to perform
this association, we remind the construction of the pentagrid using the Fibonacci tree which
is in bijection with the tiles contained in a quarter, see Figure~\ref{penta}.

\setbox110=\hbox{\epsfig{file=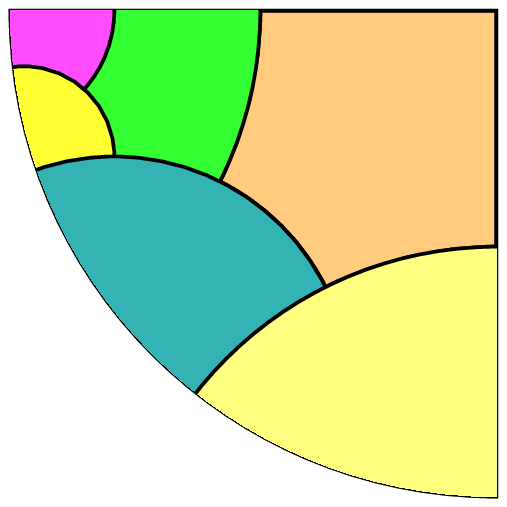,width=100pt}}
\setbox112=\hbox{\epsfig{file=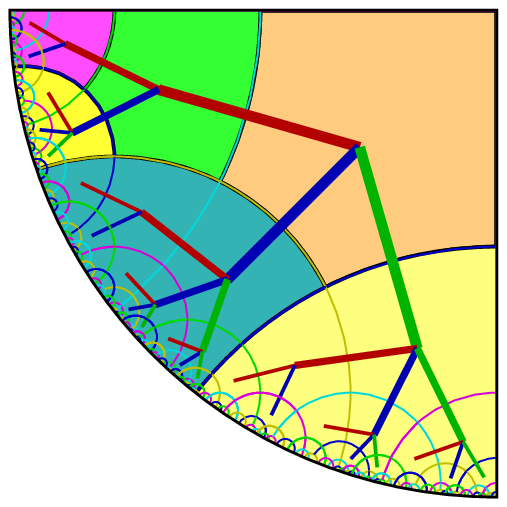,width=100pt}}
\setbox114=\hbox{\epsfig{file=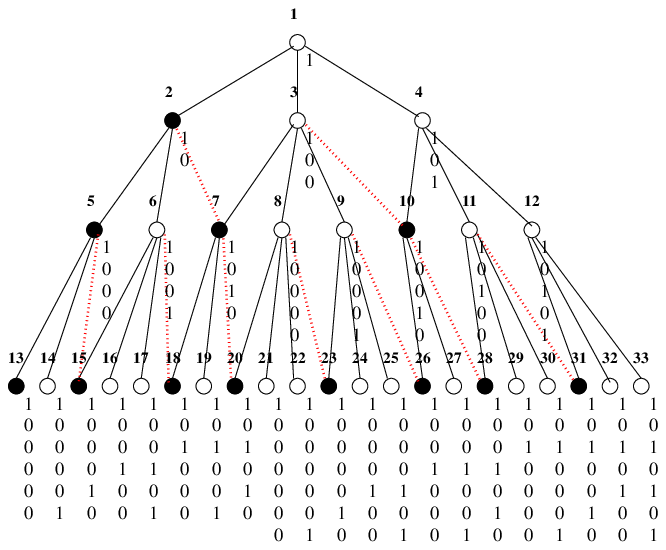,width=140pt}}
\vtop{
%\vspace{-15pt}
\ligne{\hfill
\PlacerEn {-355pt} {0pt} \box110
\PlacerEn {-255pt} {0pt} \box112
\PlacerEn {-150pt} {0pt} \box114
}
%\vspace{-35pt}
\begin{fig}\label{penta}
\leurre
Left-hand side: splitting a sector of the pentagrid. Middle: the construction of the tree
generated by the splitting. Right-hand side: the tree, called {\bf Fibonacci tree}
with its black nodes, $2$ sons, and its white ones, $3$ sons.
\end{fig}
}

   The splitting is summarized by the left-hand side picture of the figure, the middle
one showing the recursive construction of the tree yielded by the splitting: the Fibonacci
tree. The right-hand side represents the Fibonacci tree which can be defined for itself:
it has black nodes and white ones. By definition, black nodes have two sons, white ones
have three sons. For each node, exactly one of its sons is black, always the leftmost one.
The left-hand side picture of Figure~\ref{correspondence} shows how the pentagrid can
be covered by five sectors around a central cell.
\vskip 5pt
\vskip 5pt
\ligne{\hfill}
\vspace{-10pt}
\setbox110=\hbox{\epsfig{file=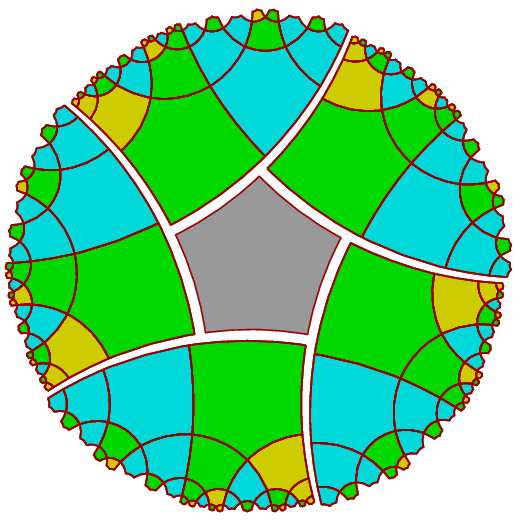,width=180pt}}
\setbox112=\hbox{\epsfig{file=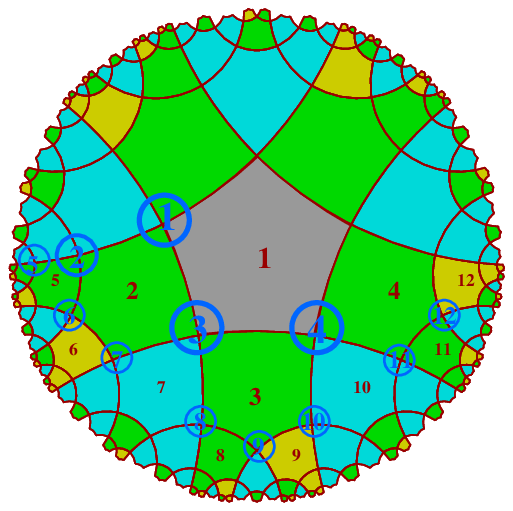,width=180pt}}
\vtop{
%\vspace{-15pt}
\ligne{\hfill
\PlacerEn {-350pt} {0pt} \box110
\PlacerEn {-170pt} {0pt} \box112
}
%\vspace{-35pt}
\begin{fig}\label{correspondence}
\leurre
The correspondence between the numbers of the pentagons and of their vertices.
\end{fig}
}
\vskip 5pt
\vskip 5pt
 Now, we can perform the announced association as follows. We start from the
splitting given by the left-hand side picture of Figure~\ref{correspondence}. On 
the right-hand side picture of the figure, we have the correspondence for the first
three levels of the pentagons. The idea is to cover all vertices of the pentagons
of a sector by numbers in a bijective way, the vertices on the rightmost branch being
excluded. Number the sides of a pentagon from~1 to~5, counter
clockwise turning around the tile, number~1 being given to the side shared by the
father of the node. We decide that the father of the head of a sector is the central
cell. With this convention, the number of a tile is given to its vertex shared by 
sides~1 and~2 for a white node and to its vertex shared by sides~2 and~3 for a black
node. It is easy to see, by induction on the level of the Fibonacci tree, that
with this process, we completely cover the vertices of all pentagons of a sector,
excepted the vertices which are on the right-hand side ray which delimit the sector,
including the vertex of the sector.

   It is not difficult to see that this algorithm can be generalized to all tilings
of the form $\{4,p\}$ for which the numbering used for $\{p,4\}$, very close to that
of the pentagrid, could be used in place of the one devised in Section~\ref{basic}.

\section*{Conclusion}

   It is interesting that both the splittings we defined in Subsections~\ref{split1}
and~\ref{split2} give rise to languages which are always regular, except for the tiling 
$\{4,5\}$ in both cases. As shown in~\cite{mmbook1}, these languages can be very different. 
In particular, there are sharp differences in the forbidden patterns for the language 
when the constant coefficient of the polynomial of the splitting is positive and when 
it is negative.

   It seems to me that each splitting has its own merit. 

   For the splitting of
Subsection~\ref{split1}, we have a polynomial of degree~3, but it is still Pisot
in all cases but one: when $p=4$ and $q=5$. The advantage of this splitting is that we 
have less tiles generated at each node by the splitting. So that the process is 
here closer to the process defined by the recursive application of the reflection in the 
sides of the previous generation. 

   The splitting of Subsection~\ref{split2} has two regions only, giving rise to a 
polynomial of degree~2 which is usually easier to solve than a cubic one. Also,
in this case, the language of the splitting is regular in all cases but one: again the
case when $p=4$ and $q=5$ as with the other splitting. However, the number 
of tiles which are generated is bigger at each generation. Moreover, in one generation 
of the splitting process we get tiles which are not in contact, neither by an edge nor 
by a vertex, with the tiles of the current generation. This is a difference with the 
traditional definition of a generation. Nevertheless, it can give a formal support to the 
notion of influence not only on the immediate neighbourhood which may have its own,
as the action at a distance in quantum physics or, in mechanics, with gravity.
And so, this splitting might have more advantages than the other.
 
   At last, the algorithm of Subsection~\ref{til45} and its generalization to
the tilings $\{4,p\}$ raises the question whether further generalizations are possible or
not. This is a question for future research.
%   It would not be suprising that the above algorithm work in all cases in the following
%way: the numbering of the tiling $\{p,q\}$ can be transported onto the tiles
%of its dual tiling $\{q,p\}$.

\end{document}